# Greedy Set Cover Estimations

## H. Aslanyan

**Yerevan State University**
1 A.Manoukyan, 375049, Yerevan, Armenia

***Abstract***. *More precise estimation of the greedy algorithm complexity for a special case of the set cover problem is given in this paper.*

## Introduction

The greedy heuristic is the most used for optimization problems. The general approach is as follows: repeatedly, a procedure that minimizes (maximizes) the local increase of the objective function is applied. In some cases the greedy strategy guarantees the optimal solution (minimal spanning trees, the shortest path, etc.), in some others - it provides acceptable approximations (e.g. disjunctive forms and tests). Typically, the greedy algorithms use simple structures that require minimal computational resource – time and memory.

Consider the set cover problem. Given a finite set $A = \{a_1, \cdots, a_n\}$ and a family of subsets of $A$, $F = \{A_1, \cdots, A_m\}$, such that every element of $A$ belongs to at least one subset from $F$, - $F$ covers $A$. The problem is in finding a collection $C \subseteq F$ of minimal size, that covers $A$.

The set cover problem is one of the most typical NP-complete problems. It has proven that there is no constant factor approximation to this problem (unless P=NP) [2].

The problem can be represented in terms of (0,1)-matrices. The elements of $A$ correspond to the columns, and each subset from $F$ corresponds to a row of the matrix. The problem is in finding the minimal number of rows that cover at least one "1" in each column. There are known reasonable approximation greedy algorithms for this problem.

Consider the following set cover greedy algorithm: – at the first step the algorithm selects the row that contains the maximal number of "1"s (covers maximal number of elements of $A$). At the current step the row, which covers the greatest number of uncovered yet elements, has been selected. It is clear that continuing this process (at most till the last row selection) all elements will be covered. Probably it may occur before. We consider the scheme [1] where a part of elements of $A$ has been covered by the greedy steps, and then, each uncovered element is being covered by taking some new row having "1" in that column. The problem is in estimating the number of all selected rows (the size of cover). We consider the estimation given for a special case in [1], and give a more precise formula for this case.

## Formula Improvement

Given a (0,1)-matrix of size $m \times n$ ($m$ is the number of rows). Each column contains at least $\gamma m$ "1"s (special case). The problem is to find minimal number of rows that contain at least one "1" on each column. Then the number of "1"s of the whole matrix is not less than $\gamma mn$, and there is a row with at least $\gamma n$ "1"s. At the first step the greedy algorithm selects the row with maximal number of "1"s, therefore the selected row will cover at least $\gamma n$ elements. Let we have done $k$ similar steps, and let the number of uncovered yet elements does not exceed $\delta_k n$. The estimate $n\delta_{k+1} \leq n\delta_k(1-\gamma)$ is the main related result, obtained in [1], Part 3, Chapter 3.5, p. 136-137. The table below outlines the part of rows, selected during the first $k$ greedy steps. The shaded columns do not contain "1"s and hence the corresponding elements are not covered yet.

|       | $a_1$ | $a_2$ | ... | ... | $a_{n(1-\delta_k)}$ | $a_{n(1-\delta_k)+1}$ | ... | ... | ... | ... | $a_n$ |
|-------|-------|-------|-----|-----|---------------------|-----------------------|-----|-----|-----|-----|-------|
| $A_1$ |       |       |     |     |                     |                       |     |     |     |     |       |
| $A_2$ |       |       |     |     |                     |                       |     |     |     |     |       |
| ...   |       |       |     |     |                     |                       |     |     |     |     |       |
| $A_k$ |       |       |     |     |                     |                       |     |     |     |     |       |
| $A_{k+1}$ |   |       |     |     |                     |                       |     |     |     |     |       |
|       |       |       |     |     |                     |                       |     |     |     |     |       |
|       |       |       |     |     |                     |                       |     |     |     |     |       |
| $A_m$ |       |       |     |     |                     |                       |     |     |     |     |       |

The summary number of "1"'s in rows, from $A_{k+1}$ to $A_m$ and part of uncovered elements (columns) is at least $\gamma m \delta_k n$. The formula improvement takes into account, that this "1"s can't be situated in the shaded part of the table. Hence, among the subsets $A_{k+1}$ to $A_m$ there exists one with at least $\dfrac{\gamma m \delta_k n}{m-k}$ "1"s. Therefore,

$$n\delta_{k+1} \leq n\delta_k - \frac{\gamma m n \delta_k}{m-k} = n\delta_k\left(1 - \frac{\gamma m}{m-k}\right) = n\delta_k(1-\gamma)\frac{m-\gamma m - k}{(m-k)(1-\gamma)} = n\delta_k(1-\gamma)\frac{1 - \frac{k}{m(1-\gamma)}}{1 - \frac{k}{m}}.$$

Using recurrently the above relation we get: $\delta_{k+1} \leq (1-\gamma)^k \dfrac{\prod_{i=1}^{k}\left(1 - \dfrac{i}{m(1-\gamma)}\right)}{\prod_{i=1}^{k}\left(1 - \dfrac{i}{m}\right)}$, which is the proposed improvement of the set cover complexity estimate in the above given special case of the problem. The additional coefficient is easy to maintain using the following formulae:

$$\left(1 - \frac{x}{y}\right)^{\frac{x-1}{2}} \leq \prod_{i=1}^{x-1}\left(1 - \frac{i}{y}\right) \leq \left(1 - \frac{x}{2y}\right)^{x-1}$$ for $x \leq y$ positive integers ($\gamma$ is supposed justified to obey this). As a result it follows that in asymptotes, when $n \to \infty$ and $m \to \infty$, the additional term is close to 1 when $\gamma \to 0$, but it is $\ll 1$ otherwise, providing the sensitive improvement of the set cover estimate.